\documentclass[5p]{elsarticle}
\usepackage{graphicx}
\usepackage{bm}
\usepackage{amsmath}
\newcommand{\subscript}[1]{\ensuremath{_{\textrm{\footnotesize{#1}}}}}

\begin{document}

\title{Applying Monte Carlo configuration interaction to transition metal dimers: exploring the balance between static and dynamic correlation}

\author[hw]{J. P. Coe}
\author[hw]{P. Murphy}
\author[hw]{M. J. Paterson}
\ead{m.j.paterson@hw.ac.uk}

\address[hw]{Institute of Chemical Sciences, School of Engineering and Physical Sciences, Heriot-Watt University, Edinburgh EH14 4AS, United Kingdom.}

\begin{abstract}
We calculate potential curves for transition metal dimers using Monte Carlo configuration interaction (MCCI).  These results, and their associated spectroscopic values, are compared with experimental and computational studies.  The multireference nature of the MCCI wavefunction is quantified and we estimate the important orbitals.  We initially consider the ground state of the chromium dimer. Next we calculate potential curves for Sc\subscript{2} where we contrast the lowest triplet and quintet states.  We look at the molybdenum dimer where we compare non-relativistic results with the partial inclusion of relativistic effects via effective core potentials, and report results for scandium nickel.     
\end{abstract}

\maketitle

\section{Introduction}
Transition metals can be very challenging for computational chemistry methods due to the existence of many important configurations at numerous geometries.  This means that methods that can cope with multireference systems, and their associated computational costs, are often necessary.  In addition, accurate modelling of heavy transition metals may also require relativistic effects to be incorporated. 

Although full configuration interaction (FCI) will produce the most accurate wavefunction in a given basis, the number of configurations that would need to be considered for transition metal dimers makes this method currently computationally intractable. A powerful approach to attempt to overcome this problem is complete active space SCF (CASSCF) \cite{siegbahn:2384} which efficiently calculates a wavefunction comprising all configurations formed from substitutions within a restricted set of orbitals by optimising coefficients and orbitals. This method is thought to model much of the static correlation which is associated with a few important configurations in the FCI wavefunction.  The remaining difference is then termed dynamic correlation and this can be accounted for by multireference CI (MRCI) or perturbative corrections such as CASPT2 \cite{CASPT2}.        

Monte Carlo configuration interaction (MCCI) \cite{mcciGreer95} aims to find a compact, yet sufficiently accurate, description of the FCI wavefunction by iteratively building up a wavefunction through the random addition of configurations without requiring knowledge of the important orbitals before a calculation.  Recent successful applications of MCCI include excitation energies \cite{GreerMcciSpectra,2013saMCCI} and properties \cite{MCCIdipoles}. MCCI with complex wavefunctions has been used in the modelling of a tunnel junction consisting of gold atoms \cite{GreerComplexMCCI}. MCCI has been demonstrated to produce potential curves of ground states \cite{MCCIpotentials} that generally compared favourably with the FCI results using just a small fraction of the FCI space for small molecules.  Excited potential curves, including avoided crossings and conical intersections, for small systems were seen to be similar to FCI results, but used a compact wavefunction when calculated with SA-MCCI in Ref.~\cite{2013saMCCI}. It is therefore interesting to investigate the potential curves of transition metal dimers with MCCI as a stern test of the method when the division between static and dynamic correlation becomes blurred and for the possibility of supporting the choice of active spaces as appropriate in other computational work.  We aim to demonstrate that MCCI can model a selection of transition metal dimers without prior knowledge of the important orbitals or resorting to perturbative corrections.  The compact nature of the MCCI wavefunction then allows us to quantify its multireference character using all of the included configurations. We also estimate the important orbitals for a representative sample of the considered geometries.  We compare the MCCI calculations with other computational work and experimental results for the chromium dimer with a cc-pVDZ and cc-pVTZ basis then Sc\subscript{2} with the same bases. We then look at the potential curve for the molybdenum dimer using a minimal basis (STO-3G) followed by LANL2DZ and Stuttgart RSC to approximate relativistic corrections. Finally we consider ScNi, with the STO-3G basis, as an example of a heteronuclear diatomic.

\section{Method}

 We use version four of the MCCI program \cite{mcciGreer95,mccicodeGreer}.  We always use the Hartree-Fock (HF) molecular orbitals (MOs) for consistency and the necessary integrals are calculated using Molpro \cite{MOLPRO_brief2012}.  The MOs are employed in the following calculations which begin from the configuration state function (CSF) formed from the occupied HF MOs. The current set of CSFs is augmented with new CSFs created using symmetry-preserving random single and double substitutions of MOs.  The Hamiltonian matrix is created and diagonalized in this set of CSFs and any new configurations with an absolute coefficient less than $c_{\text{min}}$ in the resulting wavefunction are removed.  The process is repeated and every ten iterations all CSFs in the wavefunction are considered for removal (full pruning).  The program is run until the convergence criterion for the energy, as detailed in Ref.~\cite{GreerMcciSpectra}, is satisfied.  We consider the average energy for the last three full pruning steps.  Results were calculated using an initial $10^{-3}$ Hartree convergence check and then a $5\times10^{-4}$ Hartree convergence check unless otherwise stated. We implement calculations initially at $c_{\text{min}}=5\times10^{-4}$ then lower this value until a sufficiently smooth potential curve is achieved.

We calculate vibrational energy levels of the potential curve for the most abundant isotopes with the program LEVEL 8.0 \cite{Level8}. The transition metal potential curves can display a secondary well and will not adhere to the form of the Morse potential \cite{Morse}.  We therefore use the lowest energy level to approximate the harmonic vibrational frequency ($\omega_{e}$). Due to MCCI not being size consistent, we calculate dissociation energies as the difference between the energy at equilibrium, with the ground state vibrational energy included, and the longest bond length considered. 

We quantify the multireference character associated with the MCCI wavefunction for a given basis and set of MOs using 
\begin{equation}
MR=\sum_{i} |c_{i}|^{2} -|c_{i}|^{4}.
\end{equation}
Here the $c_{i}$ are normalised so that $\sum_{i} |c_{i}|^{2}=1$. The expression is approximate when using CSFs due to their non-orthogonality. A value of zero will result from a single configuration in the wavefunction while unity will be approached as the coefficients become equal in magnitude and more numerous.  We also measure the importance of MOs in the MCCI wavefunction by considering the percentage of configurations they occur in where each occurrence is weighted by $|c_{i}|^{2}$.

\section{Results}

\subsection{Chromium dimer}
 The availability of a potential curve based on experimental results and the difficulty of the system to model makes Cr\subscript{2} useful for testing methods.  The fitted curve  \cite{Cr2ExperimentalRKR} did not display the barrier necessary for a double well, as was seen in an early computational study \cite{PhysRevLett.54.661}, but due to a lack of vibrational data the experimental potential is not known as accurately in this region.  The bond length has been found experimentally as $R_{e}=1.6788$ \r{A} \cite{Cr2ExperimentalBond} and the dissociation energy has been measured as $1.53\pm 0.06$ eV in Ref.~\cite{Mo2andCr2D0} while Ref.~\cite{Cr2ExperimentalRKR} finds  $\omega_{e}=480.6$ $cm^{-1}$ and $\omega_{e}\chi_{e}=14.1$ $cm^{-1}$.  For theoretical results, it appears that a good treatment of both dynamic and static correlation is necessary to give a sufficiently accurate potential energy curve: Ref.~\cite{RoosRevwithCr2_1998} depicts how the potential curve from CASSCF(12,12) is seen to be repulsive for small bond lengths with a weakly attractive minimum around $3$ \r{A}, but CASPT2 can produce a realistic potential.  CASPT2 and similar approaches have also been used in, e.g., Ref.~\cite{CASPT2CR22011,kurashigeCr2_2011,GVVP2CR2_2012,Matxain13} with generally high accuracy. Ref.~\cite{CASPT2CR22011} found the curve was dependent on the choice of zeroth-order Hamiltonian, but is particularly close to experimental results when using CASPT2 IPEA $0.45$ with the equilibrium bond lengths found to be around $1.68$ \r{A} and $D_{0}=1.50$ eV with a vibrational interval of $542$ $cm^{-1}$. However CASPT2 results in \cite{Matxain13} were not so accurate and found a longer bond length of  $2.43$ \r{A} and a lower dissociation energy of $1.0$ eV. Multireference coupled cluster \cite{Cr2MultiCCMuller2009} gave complete basis set estimates of $R_{e}=1.685$ \r{A}, $D_{0}=1.327$ eV and $\omega_{e}=459$ $cm^{-1}$. A double well was observed when using a TZP basis but not with a QZP basis.

When using MCCI to model an $A_{g}$ singlet in Cr\subscript{2}, we freeze 18 orbitals and, for consistency in the HF curve, we require that the HF wavefunctions have the same occupation of orbitals from each symmetry class.  The HF energies at $1.4$ \r{A} were $-2085.9767$ Hartree for cc-pVDZ and $-2085.9945$ Hartree for cc-pVTZ.  With the cc-pVDZ basis and $c_{\text{min}}=5\times10^{-4}$, the energy at bond lengths of $6$ and $3.5$ \r{A} appeared as though they may be anomalously high when using a convergence check of $10^{-3}$ Hartree. We attempted to improve the energies at these points by using the result of the next shorter bond length as a starting point. The energy at  $R=6$ \r{A} was improved substantially using this approach but the $R=3.5$ \r{A} calculation did not give a lower energy. A small barrier remains in this region when using a tighter convergence check suggesting that it may be a feature of the MCCI calculation at this c\subscript{min}.  We then considered a lower cut-off of $c_{\text{min}}=2\times10^{-4}$.  Here one point was not calculated due to time constraints.  The results were closer to the experimental curve, but a secondary well was now seen.

\begin{figure}[ht]\centering
\includegraphics[width=.4\textwidth]{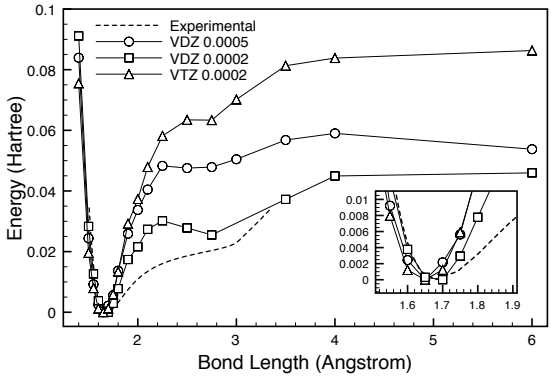}
\caption{  Energy (Hartree) shifted so that all minima are at zero against bond length (\r{A}) for the chromium dimer for the experimental results compared with MCCI ($c_{\text{min}}=5\times10^{-4}$ or $c_{\text{min}}=2\times10^{-4}$) using cc-pVDZ and MCCI ($c_{\text{min}}=2\times10^{-4}$) using cc-pVTZ both with 18 frozen orbitals. Inset: An enlarged view of the minima.}\label{fig:Cr2curve}
\end{figure}

For the cc-pVTZ basis we initially used $c_{\text{min}}=5\times10^{-4}$ however this resulted in a curve which was not smooth.  Hence the points were recalculated using $c_{\text{min}}=2\times10^{-4}$ with the larger c\subscript{min} result used as the starting point.  This gave a smooth curve except for two points which appeared too high in energy compared with the rest of the curve: $2.75$ and $6$ \r{A}.  They were much improved when the calculation was restarted using the configurations of the next shorter bond length as the starting point.  The results are displayed in Fig.~\ref{fig:Cr2curve} with the fitted experimental results from Ref.~\cite{Cr2ExperimentalRKR} also plotted for comparison.  The results agree with the experimental curve for the inner wall and equilibrium geometry, but at intermediate geometries the cc-pVDZ curve increases too rapidly and this effect is more pronounced when using the cc-pVTZ basis although this increase is moderated when using cc-pVDZ with the lower c\subscript{min}.  We note that CASPT2 results can be very close to the experimental curve but this requires tailoring of both the active space and zeroth-order Hamiltonian.

On lowering the cut-off in cc-pVDZ the vibrational frequency changed from $734$ to $490$ $cm^{-1}$ giving a better agreement with experiment, but the dissociation energy reduced from $1.41$ to $1.22$ eV. The equilibrium bond length changed from approximately $1.65$ to $1.70$ \r{A}.   
For the cc-pVTZ results we find that the vibrational frequency is $530$ $cm^{-1}$ and $D_{0}=2.32$ eV. We see that the larger basis  gives a dissociation energy which is too high compared with the experimental result of $1.53$ eV \cite{Mo2andCr2D0} and other computational work. This suggests that the MCCI curve is improved around the equilibrium more  than when approaching dissociation for this c\subscript{min} when increasing the basis size to cc-pVTZ.  The equilibrium bond length was approximately $1.65$ \r{A} when using cc-pVTZ.  This geometry is in reasonable agreement with the experimental result given the number of points considered.  

\begin{figure}[ht]\centering
\includegraphics[width=.4\textwidth]{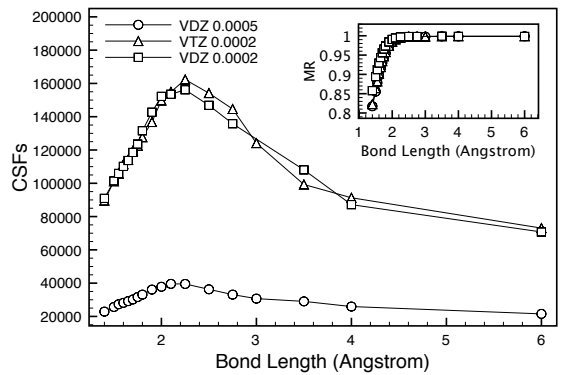}
\caption{Number of CSFs at convergence for the MCCI wavefunction against bond length (\r{A}) for the chromium dimer when using the cc-pVDZ basis or the cc-pVTZ basis at $c_{\text{min}}=2\times10^{-4}$ or $c_{\text{min}}=5\times10^{-4}$. Inset: A measure of the multireference character ($MR$) against bond length (\r{A}) }\label{fig:Cr2csfs}
\end{figure}

The FCI space for cc-pVDZ is around $10^{15}$ SDs when freezing 18 orbitals and taking symmetry into account while MCCI used approximately $3.1\times10^{4}$ CSFs on average with the larger cutoff and $1.2\times 10^{5}$ with the smaller cutoff.  For cc-pVTZ there are $10^{18}$ SDs in the FCI space and $1.2\times10^{5}$ CSFs in the MCCI wavefunction on average.  The variation in the size of the MCCI wavefunction with geometry is depicted in Fig.~\ref{fig:Cr2csfs} where the 
smooth curve suggests that the method is treating different geometries reasonably consistently as there are no single points which stand out as having been fortuitously calculated to different accuracy within MCCI. When considering the number of CSFs, the system appears to be most challenging at intermediate geometries a little longer than equilibrium. The multireference character for the MCCI wavefunction with these MOs is high all along the curve (inset of Fig.~\ref{fig:Cr2csfs}) and increases with bond length to essentially saturate on the scale of the graph despite the number of configurations falling as dissociation is approached.  The very large FCI space for cc-pVTZ and strong multireference character may be partly responsible for the large dissociation energy as a smaller fraction of the FCI space is considered compared with cc-pVDZ at either cut-off despite using wavefunctions containing up to $1.6\times 10^{5}$ CSFs. 

We extrapolate the $c_{\text{min}}=2\times10^{-4}$ results to approximate the complete basis set limit (CBS). The scheme of Ref.~\cite{KartonMartin} is used for the CBS limit for the Hartree Fock energy  where $E_{x}=E_{\infty}+A(x+1)e^{-9\sqrt{x}}$.  Here $x=2$ for cc-pVDZ and $x=3$ for cc-pVTZ.   The MCCI correlation energy is extrapolated through the use of $E_{corr,x}=E_{corr,\infty}+Bx^{-3}$ \cite{HelgakerScheme}. We find that the estimate does not change the equilibrium geometry at cc-pVTZ, however $\omega_{e}=488$ $cm^{-1}$ and the overestimate in the binding energy when we used MCCI with cc-pVTZ is propagated to the CBS estimate of  $D_{0}=2.75$ eV.  We acknowledge that the cc-pVDZ and cc-pVTZ are still reasonably far from the basis set limit and that MCCI did not appear so successful at the cut-off used when faced with the very large configuration space from using the cc-pVTZ basis so the estimated CBS results should be treated with caution.

Defining an important MO as one whose weighted importance is greater than $10\%$ for alpha or beta spin, we find eight such MOs at $R=1.4$ \r{A} while this rises to twelve for the longest bond length
when using cc-pVTZ. The six non-frozen reference MOs appear to be important at all geometries sampled and there are 15 important orbitals in total.  For cc-pVDZ with $c_{\text{min}}=2\times10^{-4}$ the reference MOs also appear in all sampled geometries and there are 10 important orbitals at $R=1.4$ \r{A} and 12 at $R=1.6$ \r{A}.

\subsection{Scandium dimer}

Despite Sc\subscript{2}  being the subject of many studies its experimental bond length has not been determined and theoretical dissociation energies \cite{Witek2010Sc2,ScXdftResults,GVVP2CR2_2012,KaplanMn2andSc2MRCISD} often appear a little high compared with experimental results. For example, GVVPT2 results \cite{GVVP2CR2_2012} find  $D_{0}=2.36$ eV,  $\omega_{e}=225.9$  $cm^{-1}$ and $R_{e}=2.57$ \r{A} when using cc-pVTZ with the quintet state lower by $0.23$ eV.  We use the experimental dissociation energy of $1.12\pm 0.22$ eV from Ref.~\cite{Sc2D0exp}. For the vibrational energy levels the experimental results \cite{RamanMn2andSc2} are $\omega_{e}=239.9$ $cm^{-1}$ and $\omega_{e}\chi_{e}=0.93$ $cm^{-1}$ for the Morse curve.   Experimentally the ground state has been given as a quintet \cite{GroundStateExpSc2}.  However the lowest lying triplet $^{3}\Sigma^{-}_{u}$ and quintet state  $^{5}\Sigma^{-}_{u}$ appear to be close in energy and although most theoretical work has found the quintet to be the ground state \cite{Witek2010Sc2,GVVP2CR2_2012,KaplanMn2andSc2MRCISD} the results are not unanimous \cite{MatxainTripletSc2}.  

\begin{figure}[ht]\centering
\includegraphics[width=.4\textwidth]{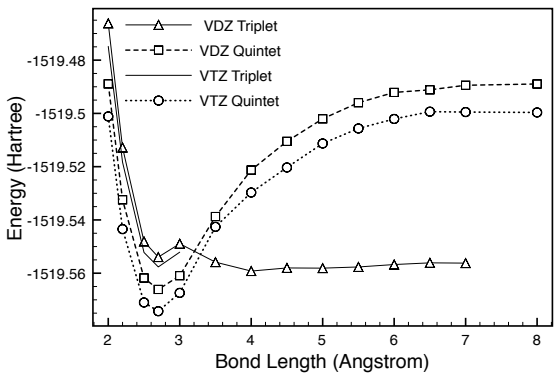}
\caption{Sc\subscript{2} triplet and quintet $A_{u}$ states using the cc-pVDZ or cc-pVTZ basis, $c_{\text{min}}=2\times10^{-4}$ and 17 frozen orbitals.}\label{fig:Sc2}
\end{figure}

It has been noted, in Ref.~\cite{C1CP20177H} and references therein, that if the active space is not large enough then the ground state could be a triplet depending on the method used to deal with intruder configurations. The susceptibility of the spin of the ground state to the choice of active space, in addition to the challenge of this being a multireference system, make this system a particularly interesting candidate for MCCI calculations. We calculate the triplet and quintet curves of $A_{u}$ symmetry in $D_{2h}$ which will contain the $\Sigma^{-}_{u}$ state.  The HF energies at $2$ \r{A} when using cc-pVDZ are $-1519.3286$ Hartree for the quintet and $-1518.0161$ Hartree for the triplet while the values for cc-pVTZ are -$1519.3327$, and $-1518.02451$ Hartree.  We freeze $17$ orbitals and find that for $c_{\text{min}}=2\times10^{-4}$ sufficiently smooth curves could be observed as depicted in Fig.~\ref{fig:Sc2}. Here the final convergence criterion was $10^{-3}$ for cc-pVDZ and $5\times10^{-4}$ for cc-pVTZ.

 For the cc-pVDZ results we see that the quintet is the lower state by $0.33$ eV  at equilibrium and the triplet curve shape suggests a crossing between states that would be of different symmetry in the full $D_{\infty h}$ point group.  This agrees with the results of Ref.~\cite{GVVP2CR2_2012} where the $^{3}\Sigma_{u}^{-}$ and $^{3}\Sigma_{g}^{-}$ cross however in this work the curves are both of $A_{u}$ symmetry so would be expected to be $^{3}\Sigma_{u}^{-}$ and $^{3}\Delta_{u}$.  Our results indicate an equilibrium geometry of approximately $2.7$ \r{A} and we find for the ground vibrational energy level that $\omega_{e}=216$ $cm^{-1}$.  We calculate the dissociation energy as $D_{0}=2.09$ eV. The quintet MCCI results used on average  $43,497$ CSFs compared with an FCI SD space of around $4\times10^{10}$ when considering symmetries.
The FCI space for the triplet is around $7\times10^{10}$ SDs when symmetry is taken into account. The triplet results used $53,268$ CSFs on average. There is a noticeable change in the number of configurations after the suggested crossing: the results for $R\leq3$ \r{A} used $\sim 83000$ on average while those for larger R used  $\sim 35 000$.

When using cc-pVTZ the result at $R=8$ \r{A} appeared too high so we used the configurations of $R=7$ \r{A} as a starting point which resulted in a lower energy. We find that $R_{e}$ stays at approximately $2.7$ \r{A},  $\omega_{e}=222$ $cm^{-1}$ and $D_{0}=2.02$ eV.  Again the quintet is the ground state but the gap between the spin states is now $0.45$ eV. Here we calculate the triplet curve only to the implied crossing.  The quintet used 50592 CSFs on average in MCCI compared with an FCI space of $3\times10^{12}$ SDs. The triplet calculated to $R=3$ \r{A} used $94098$ CSFs on average in MCCI while the FCI space was around $6\times10^{12}$ SDs. The spectroscopic values from MCCI in this case are robust against an increase in the basis size and are similar to many of the other wavefunction based computational studies with the accompanying overestimate in $D_{0}$ compared with experiment. 

\begin{figure}[ht]\centering
\includegraphics[width=.4\textwidth]{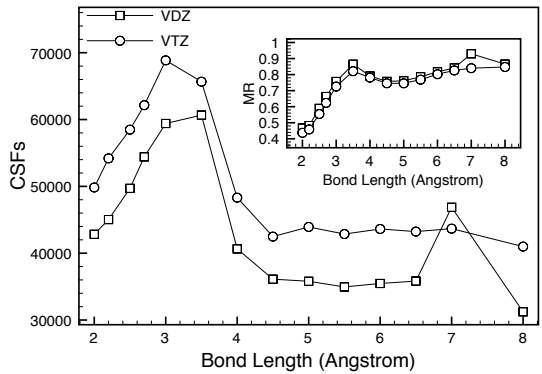}
\caption{Number of CSFs at convergence against bond length (\r{A}) for the scandium dimer using the cc-pVDZ or cc-pVTZ basis. Inset: A measure of multireference character versus bond length (\r{A}).}\label{fig:Sc2csfs}
\end{figure}

We see in Fig.~\ref{fig:Sc2csfs} that the number of configurations peaks at intermediate geometries similarly to the Cr\subscript{2} results. Compared with Cr\subscript{2} the system is less strongly multireference and the multireference character has a local maximum at an intermediate geometry.  The number of CSFs is not as large in a triple-zeta basis compared with Cr\subscript{2} which may be attributed to the smaller FCI space and slightly lower multireference character.  The number of CSFs and multireference measure suggest that the result for $R=7$ \r{A} in cc-pVDZ may have been calculated less accurately and the potential energy curve appears to have a small increase at this point (Fig.~\ref{fig:Sc2}). However the curves do not suggest any problems with the cc-pVTZ results which is perhaps due to the tighter convergence criterion here.  By extrapolating the cc-pVDZ and cc-pVTZ results for the quintet state to approximate the CBS limit we find that the minimum remains at $2.7$ \r{A} while $\omega_{e}=195$ $cm^{-1}$ and $D_{0}=1.99$ eV.

When using cc-pVTZ, the $6$ MOs from the reference wavefunction are found to be important at all sampled bond lengths and are the only important MOs at the considered shortest bond lengths. This with the MR results suggest that the correlation may be thought of as mainly dynamic at short bond lengths when using these MOs. The  $9$ important MOs for the longest bond length used are the overall important orbitals.

\subsection{Molybdenum dimer}

Molybdenum, with atomic number $42$, is the heaviest element that we consider in this work.  Relativistic effects may therefore be expected to be more important.  We compare a non-relativistic result using the STO-3G basis with the approximation of relativistic effects via an effective core potential (ECP) basis: the LANL2DZ basis \cite{LANL2DZ}.

Experimentally, the equilibrium bond length has been found \cite{Mo2ReExp} to be $1.94$ \r{A}. Ref.~\cite{Efremov78Mo2} finds the experimental bond length as $R_{e}=1.929$ \r{A} while $\omega_{e} = 477.1$ $cm^{-1}$ and $D_{0}=4.14\pm0.65$ eV. A more recent study \cite{Mo2andCr2D0} gives the dissociation energy as $4.474\pm0.01$ eV.

Computational studies \cite{PhysRevLett.54.661,Mo2Zhu2002,Mo2Borin2007} have often produced relatively good agreement with these spectroscopic constants. For example CASSCF/MS-CASPT2 with a (12,12) active space and a quadruple-zeta ANO-RCC basis set gave $R_{e}=1.950$ \r{A}, $\omega_{e}=459$ $cm^{-1}$ and $D_{0}=4.41$ eV \cite{Mo2Borin2007}. However Ref.~\cite{Matxain13} found $D_{0}=2.14$ eV with CASPT2 using an ECP basis.

\begin{figure}[ht]\centering
\includegraphics[width=.4\textwidth]{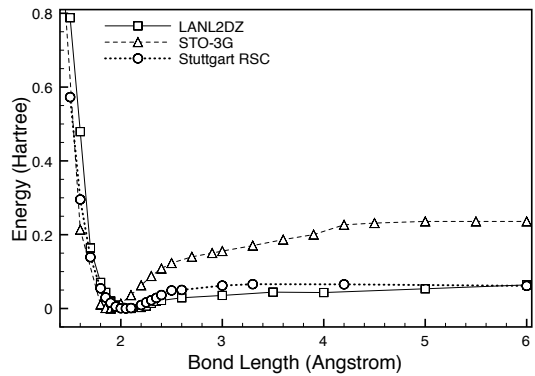}
\caption{MCCI energies, shifted so the minima are at zero, for the molybdenum dimer with $c_{\text{min}}=5\times10^{-4}$ against bond length (\r{A}) using the STO-3G, LANL2DZ or Stuttgart RSC 1997 ECP basis all with 12 non-frozen electrons.}\label{fig:Mo2LANL2DZ}
\end{figure}

We calculate the singlet of $A_{g}$ symmetry in $D_{2h}$ using MCCI with $c_{\text{min}}=5\times10^{-4}$.  The HF occupancy was fixed to give a smooth curve for the STO-3G basis, where, at a bond length of $1.5$ \r{A}, the HF energy was $-7870.9969$ Hartree. The MCCI results using these MOs with 36 orbitals frozen are shown in Fig.~\ref{fig:Mo2LANL2DZ}.  The MCCI curve is smooth when the convergence criterion is lowered to $5\times10^{-4}$ Hartree and we find that approximately $R_{e}=1.9$ \r{A}, $\omega_{e}=467$ $cm^{-1}$ and  $D_{0}=6.4$ eV.  The MCCI results used, on average, 13258 CSFs while the FCI space would be around $4\times10^{7}$ SDs when symmetry is taken into account. The dissociation energy is too high, but the other results are in agreement with experiment and much other computational work. However the minimal basis means that any comparisons should be made cautiously.

 When using the LANL2DZ basis, 56 electrons are taken into account by the ECP and we then freeze a further 8 orbitals.  Here the HF energy at $R=1.5$ \r{A} was $-132.7382$ Hartree.  With a convergence check of $10^{-3}$, preliminary results had a sharp jump in energy at $R=2.3$ \r{A} so we use a fixed HF occupancy from $R=2.2$ \r{A} onwards and the configurations of the previous smaller geometry as a starting point for the MCCI calculation. This produced a smooth curve except the energy at $R=4$ \r{A} appeared too high. By allowing Molpro \cite{MOLPRO_brief2012} to guess the HF initial occupancy at this point we find that the MCCI result is in line with the rest of the curve. We then continued the calculations but with a $5\times10^{-4}$ convergence check and the results are displayed in Fig.~\ref{fig:Mo2LANL2DZ}. The LANL2DZ results with $c_{\text{min}}=5\times10^{-4}$ suggest that the equilibrium geometry is approximately $2.1$ \r{A}. We find that $\omega_{e}=253$ $cm^{-1}$ and $D_{0}=1.73$ eV.  This used $27464$ CSFs on average compared with an FCI space of the order of $10^{11}$ SDs when symmetry is taken into account.

The LANL2DZ curve is less strongly binding than the experimental results and other computational work.  We note that the LANL2DZ basis gave a very low dissociation energy and excessive bond length for the chromium dimer using multireference perturbation in Ref.~\cite{GVVP2CR2_2012}.   The shape suggests that there may be a crossing of two curves of different symmetry within the full $D_{\infty h}$ point group.  Hence we also continued the calculations at certain points with SA-MCCI \cite{2013saMCCI} to investigate the first excited singlet state of $A_{g}$ symmetry in $D_{2h}$. However the results did not reveal a curve crossing.   

We also employed the Stuttgart RSC (relativistic small core) 1997 ECP basis as stored at the basis set exchange \cite{BasisSetExchange}. In this case the HF energy at $R=1.5$ \r{A} was $-134.2388$ Hartree and the MCCI results were more similar to those found with LANL2DZ than with STO-3G.

\begin{figure}[ht]\centering
\includegraphics[width=.4\textwidth]{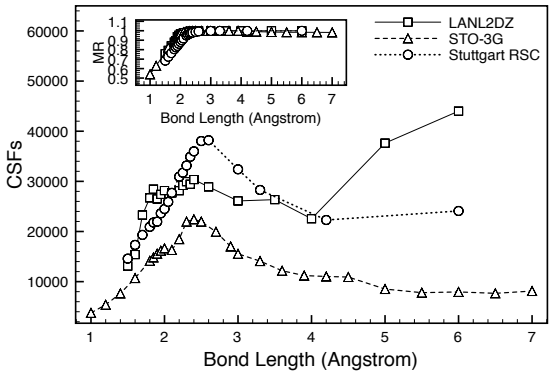}
\caption{Number of CSFs at convergence for the MCCI wavefunction with $c_{\text{min}}=5\times10^{-4}$  against bond length (\r{A}) for the molybdenum dimer with the STO-3G, LANL2DZ or Stuttgart RSC 1997 ECP basis. Inset: A measure of multireference character versus bond length (\r{A}).}\label{fig:Mo2csfs}
\end{figure}

The number of CSFs for the MCCI wavefunction is plotted against the geometry of the system in Fig.~\ref{fig:Mo2csfs}.  When using STO-3G or the Stuttgart RSC 1997 ECP the curve is fairly smooth and peaks a little after the equilibrium bond length in line with the results from the other dimers considered in this work. The LANL2DZ basis results are less smooth but behave similarly until $R=4$ \r{A} where the number of CSFs increases sharply. There does not appear to be a problem with the energy results around this geometry though and the plot of the number of CSFs does not illuminate the possibility of a curve crossing around $R=2.2$ \r{A}.  The multireference nature generally increased with bond length for both bases although there is a very small drop as dissociation is approached (inset of Fig.~\ref{fig:Mo2csfs}) when using STO-3G.   

In the STO-3G results the $6$ reference MOs are generally important but at the longest bond length only $5$ are found to be important. There are between $1$ and $7$ other orbitals at the sampled bond lengths and  overall we find $13$ important orbitals.  For the LANL2DZ results we estimate there are $14$ important orbitals. We see that the $6$ reference MOs are classified as important across the curve. The are $9$ important MOs at the shortest bond length considered and this rises to $13$ at $R=6.0$ \r{A}.  The $6$ reference MOs appear to be important at all geometries when using the Stuttgart RSC also. In this case there are $7$ important MOs at $R=1.5$ \r{A} while $12$ are deemed important at $R=6.0$ \r{A}.

\section{Scandium nickel dimer}

ScNi has historically proven to be a difficult species to characterise both experimentally and computationally in terms of both equilibrium bond length and dissociation energy. This species therefore would appear to be a good candidate on which to test the efficacy of MCCI for heteronuclear diatomic transition metals.  The ground state of this species is routinely accepted as $^{2}\Sigma^{+}$ after agreement between computational and experimental work \cite{Zee84,Zee88,Miedema80,Arrington95,Faegri91}.
Experimentally, there are no data for either dissociation energy or equilibrium bond length. Ref.~\cite{Arrington95} does however provide details of vibrational levels with $\Delta G_{1/2}=334.5\pm 1.0$ $cm^{-1}$,  which we use to approximate $\omega_{e}$, and indicates that a dissociation energy of $D_{0}>1.36$ eV is expected.  An empirical model \cite{Miedema80} was used to predict $D_{0} = 3.28$ eV.  CASSCF calculations  \cite{Faegri91} found $R_{e} = 2.04$ \r{A} whilst MRCI+Q resulted in $R_{e} = 2.13$ \r{A}, $D_{e} = 1.55$ eV and $\omega_{e} = 322$ $cm^{-1}$.  Local-density-functional-LCAO work \cite{Mattar92} reported values of $R_{e} = 2.019$ \r{A}, $D_{e} = 5.95$ eV and $\omega_{e} =  405.9$ $cm^{-1}$. The large value of $D_{e}$ is put down to the known problem of overestimation of binding energies in the method.  Density-functional theory with the BPW91 approximation and 6-311+G* basis set found \cite{ScXdftResults} $R_{e} = 2.047$ \r{A}, $D_{e} = 3.30$ eV and $\omega_{e}= 349$ $cm^{-1}$. 

\begin{figure}[ht]\centering
\includegraphics[width=.4\textwidth]{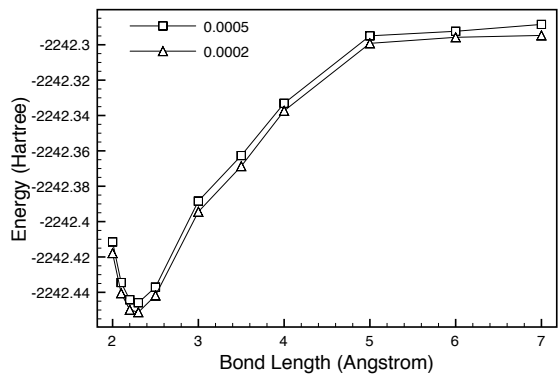}
\caption{MCCI energies for ScNi with $c_{\text{min}}=5\times10^{-4}$ and $c_{\text{min}}=2\times10^{-4}$ against bond length (\r{A}) using the STO-3G basis with 18 frozen orbitals.}\label{fig:ScNi}
\end{figure}

We calculate the doublet of $A_{1}$ symmetry in $C_{2v}$ using MCCI with $c_{\text{min}} = 5\times10^{-4}$ and $c_{\text{min}}=2\times10^{-4}$ with a convergence criterion of $10^{-3}$ Hartree.  We fix the orbital occupancy to ensure an overall wavefunction symmetry of $A_{1}$ and doublet spin. For the STO-3G basis set, we freeze 18 orbitals. A smooth HF potential energy curve is obtained by allowing calculations at bond length $3.00$ \r{A} to start from the previous result for $2.50$ \r{A} and by allowing calculations at bond length $3.50$ \r{A}  to start from the previous result for $4.00$ \r{A}. All other bond lengths required no further manipulation.  The HF energy at a bond length of $2$ \r{A} was $-2242.0030$ Hartree.  From the subsequent MCCI curve, shown in Fig.~\ref{fig:ScNi}, we report approximate results $R_{e} = 2.30$ \r{A}, $D_{0} = 4.26$ eV and $\omega_{e}=329$ $cm^{-1}$.

We then move to $c_{\text{min}} = 2\times10^{-4}$ whilst retaining the convergence criteria of $10^{-3}$ Hartree and obtain approximate results $R_{e} = 2.30$ \r{A}, $D_{0} = 4.12$ eV and $\omega_{e}= 309$ $cm^{-1}$. The equilibrium bond lengths appear to be slightly longer than value of $2.13$ \r{A} using MRCI+Q \cite{Faegri91} and the dissociation energy is somewhere in the middle of the currently published calculations. Relatively good agreement is found between the calculated and experimental values for $\omega_{e}$.  However, the use of minimal basis sets such as STO-3G means that conclusive statements cannot be made.

\begin{figure}[ht]\centering
\includegraphics[width=.4\textwidth]{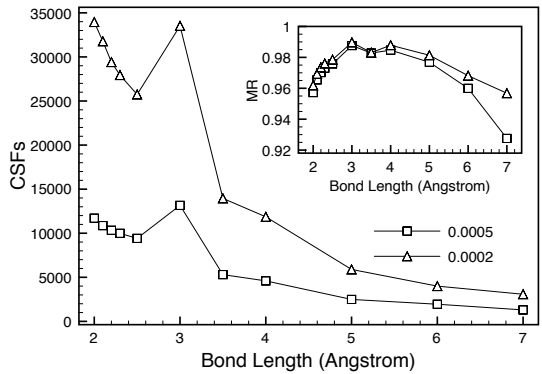}
\caption{Number of CSFs at convergence for the MCCI wavefunction with $c_{\text{min}}=5\times10^{-4}$ and $c_{\text{min}}=2\times10^{-4}$  against bond length (\r{A}) for ScNi with the STO-3G basis.  Inset: A measure of multireference character versus bond length (\r{A}).}\label{fig:ScNiCSFs}
\end{figure}

  The FCI space consists of around $10^{8}$ SDs with MCCI results using on average $7372$ CSFs for $c_{\text{min}} = 5\times10^{-4}$ and 20102 CSFs for $c_{\text{min}} = 2\times10^{-4}$.  The number of CSFs for the MCCI wavefunction is plotted in Fig.~\ref{fig:ScNiCSFs}. As can be seen a general decline in the number of CSFs is found as the bond length increases from equilibrium with a single spike at $3.00$ \r{A} in both $c_{\text{min}} = 5\times10^{-4}$ and $c_{\text{min}} = 2\times10^{-4}$ cases. When looking at the multireference nature, this coincides with a peak also at $3.00$ \r{A} as seen in the inset of Fig.~\ref{fig:ScNiCSFs}, but the two curves otherwise appear not to be correlated.  It should be noted that ScNi is highly multireference in nature across the entire potential energy surface.  Attempts to use 3-21G, cc-pVDZ and cc-pVTZ basis sets using accuracy as high as $c_{\text{min}} = 2\times10^{-4}$ with convergence threshold $5\times10^{-4}$ Hartree were unsuccessful in producing smooth MCCI curves and this is believed to be partly due to HF convergency problems and the size of the FCI space being $10^{13}$, $10^{17}$ and $10^{20}$ respectively despite freezing 18 orbitals.

 Analysis of the important orbitals, at $c_{\text{min}}=2\times10^{-4}$, indicated that the seven MOs occupied in the reference are important at all bond lengths with between three and five of the unoccupied MOs considered important. Overall we find 14 important MOs.

\section{Summary}

We have used Monte Carlo configuration interaction (MCCI) to calculate potential curves of four transition metal dimers that present a computational challenge due to their large FCI spaces and multireference character. 

For Cr\subscript{2} the MCCI potential curves were fairly similar to the experimental result at the equilibrium and shorter geometries, but seemed to increase too quickly as the bond lengthened. 
 The results when using cc-pVDZ agreed reasonably well with experiment (Table~\ref{tbl:Results}). We note that the cc-pVDZ results became closer to the experimental potential curve as c\subscript{min} was lowered but the lower dissociation energy agreed less well. However the dissociation energy became too high when using a cc-pVTZ basis even when compared with the cc-pVDZ results at the larger cutoff.  When using cc-pVTZ the dimer had the largest FCI space and smallest fraction captured by MCCI when using a reasonable cut-off (Table~\ref{tbl:FractionFCI}) and this could be responsible for what seemed to be a much better description around the equilibrium geometry than when approaching dissociation of the strongly multireference wavefunction. This suggests that approaches to reduce the size consistency error in MCCI at reasonable c\subscript{min} values may be worth investigating to improve the accuracy of dissociation energies when the FCI space is very large.

\begin{table}[h]
\centering
\caption{Summary of MCCI results at the smallest c\subscript{min} considered and experimental results.} \label{tbl:Results}
\begin{tabular*}{8.5cm}{@{\extracolsep{\fill}}lcccc}
\hline
\hline
Dimer & Basis & $R_{e}$ (\r{A}) & $\omega_{e}$ ($cm^{-1}$) & $D_{0}$ ($eV$)  \\
\hline
Cr\subscript{2}  & cc-pVDZ  & $1.70$  & $490$ & $1.22$  \\
  & cc-pVTZ & $1.65$  & $530$  & $2.32$  \\
  &Exp   & $1.6788$\cite{Cr2ExperimentalBond}  & $480.6$\cite{Cr2ExperimentalRKR}  & $1.53$\cite{Mo2andCr2D0}  \\
Sc\subscript{2}  & cc-pVDZ   & $2.7$ & $216$ & $2.09$ \\
  & cc-pVTZ   & $2.7$ & $222$ & $2.02$ \\
    & Exp   & - & $239.9$\cite{RamanMn2andSc2} & $1.12$\cite{Sc2D0exp} \\
Mo\subscript{2}  & STO-3G  & $1.9$ & $467$ & $6.40$ \\
  &  LANL2DZ   & $2.1$ & $253$ & $1.73$ \\
  & Stuttgart   & $2.05$  & $338$  & $1.65$ \\
  & Exp   & $1.94$\cite{Mo2ReExp} & $447.1$ \cite{Efremov78Mo2} & $4.474$ \cite{Mo2andCr2D0} \\
ScNi& STO-3G  & $2.3$ & $309$ & $4.12$ \\ 
  & Exp   & - & $335$ \cite{Arrington95} & $>1.36$ \cite{Arrington95} \\
\hline
\hline
\end{tabular*}
\end{table}

In agreement with most other work, the MCCI results for Sc\subscript{2} gave the quintet as the ground state rather than the triplet when using the cc-pVDZ or cc-pVTZ basis. The spectroscopic values appeared stable on increasing the size of the basis and were also in general agreement with other computational work in that $\omega_{e}$ agreed reasonably well with experiment but $D_{0}$ was too high.

\begin{table}[h]
\centering
\caption{Approximate average size of the ground-state wavefunctions at the lowest c\subscript{min} considered.} \label{tbl:FractionFCI}
\begin{tabular*}{8.5cm}{@{\extracolsep{\fill}}lcccc}
\hline
\hline
Dimer & Basis & MCCI CSFs & FCI SDs & Fraction  \\
\hline
Cr\subscript{2}  & cc-pVDZ  & $1.2\times10^{5}$  &  $10^{15}$ & $10^{-10}$  \\
  & cc-pVTZ & $1.2\times10^{5}$ & $10^{18}$ & $10^{-13}$ \\
 
Sc\subscript{2}  & cc-pVDZ   & $4.3\times10^{4}$ & $4\times10^{10}$  & $10^{-6}$  \\
  & cc-pVTZ   & $5.1\times 10^{4}$ & $3\times10^{12}$ &  $10^{-8}$ \\

Mo\subscript{2}  & STO-3G  & $1.3\times 10^{4}$  &  $4\times10^{7}$  & $10^{-4}$  \\
  &  LANL2DZ   &  $2.7\times10^{4}$ & $10^{11}$  & $10^{-7}$  \\
  &  Stuttgart   & $2.7\times10^{4}$               &  $7\times 10^{14}$          &  $10^{-11}$           \\
ScNi  & STO-3G  & $2\times 10^{4}$  &  $10^{8}$  & $10^{-4}$  \\
\hline
\hline
\end{tabular*}
\end{table}

Table~\ref{tbl:Results} shows that Mo\subscript{2} modelled with an STO-3G basis gives better results than when using LANL2DZ or Stuttgart RSC to account for some of the relativistic corrections via an effective core potential. However the dissociation energy is too high when using a minimal basis.  MCCI might be expected to give a fairly similar level of accuracy to the Sc\subscript{2} results based on the fraction of the FCI space recovered for LANL2DZ (Table~\ref{tbl:FractionFCI}) although the Mo\subscript{2} wavefunctions are more strongly multireference. This could suggest that MCCI is capturing enough of the FCI wavefunction but that the bases are not suitable for describing the potential curve of Mo\subscript{2}. With larger bases though the fraction of the FCI space recovered may be a problem as suggested for the chromium dimer.

Bases larger than STO-3G for ScNi appeared to be outside the range of applicability of the method as smooth potential curves could not be achieved.  This was suggested as a consequence of the large FCI space and HF convergence problems of this strongly multireference nature heteronuclear diatomic.  The minimal basis calculation gave a similar vibrational frequency to experiment and was not in disagreement with the predicted dissociation energy although the latter was on the high side of previous computational work. 

The multireference measure we introduced ($MR$) showed that the MCCI wavefunction was strongly multireference for all the dimers but with Sc\subscript{2} a little less so than the others.  The multireference nature tended to increase with the bond length, for all but ScNi, unlike the number of MCCI configurations which would peak at an intermediate geometry with the exception of Mo\subscript{2} with LANL2DZ.  We estimated the important molecular orbitals over a range of geometries and found that the total ranged from $9$ for Sc\subscript{2} to $15$ for Cr\subscript{2}. This may hint at an appropriate size of a fixed active space necessary for CAS calculations on these systems and bases.

One current limitation of this method is that when the potential curve is very shallow then the stochastic error can mask the shape of the curve at reasonable c\subscript{min}. We observed this problem for Mn\subscript{2} where the bonding has been demonstrated \cite{Mn2Tzeli2008} to be van der Waals with a dissociation energy of around $0.05$ eV. Hence the curve seems to be insufficiently deep for this Monte Carlo approach to be efficiently applied.  For low enough c\subscript{min} and sufficiently long calculations then the MCCI results should be insensitive to the HF orbitals for an all-electron computation.  However many orbitals need to be frozen and reasonable values of c\subscript{min} used in these systems for tractable calculations. Hence we found that the HF calculation can impact the MCCI results and the occupancy of the HF determinant may need to be fixed. Furthermore, previous MCCI results may have to be used as a starting point for similar geometries to achieve the lowest energy smooth curve by increasing the chance that important configurations have been found at all geometries when dealing with strongly multireference systems with large FCI spaces.  We have demonstrated, however, that compact wavefunctions
can be found using MCCI to describe parts of the potential energy curves of four transition metal dimers reasonably well, and with qualitatively correct general features, without the need for perturbative corrections or requiring insight in the choice of orbitals for an active space.  These wavefunctions then offer the possibility of the calculation of other properties and of perhaps constructing an active space for CAS calculations via MCCI natural orbitals \cite{MCCInatorb}.

\section*{Acknowledgments}
We thank the European Research Council (ERC) for funding under the European Union's Seventh Framework Programme (FP7/2007-2013)/ERC Grant No. 258990.

\providecommand{\noopsort}[1]{}\providecommand{\singleletter}[1]{#1}%

\end{document}